\documentclass[a4paper,10pt,twoside]{cpc-hepnp}
\usepackage{multicol}
\usepackage{graphicx}
\usepackage{booktabs}
\usepackage{amssymb,bm,mathrsfs,bbm,amscd}
\usepackage[tbtags]{amsmath}
\usepackage{lastpage}
\usepackage{url}

\begin{document}

\fancyhead[c]{\small Submitted to Chinese Physics C}
\fancyfoot[C]{\small \thepage}


\title{Study of a sealed high gas pressure THGEM detector and response of Alpha particle spectra\thanks{Supported by the National Natural Science Foundation of China (Grant No. 11575193, 11205240, 11265003, U1431109). }}

\author{%
      Yu-Ning ZHANG$^{1}$ 
\quad Qian LIU$^{1,2;1)}$\email{liuqian@ucas.ac.cn}%
\quad Hong-Bang LIU$^{3;2)}$\email{liuhb@gxu.edu.cn}%
\quad Yi-Gang XIE$^{1,4}$\\
\quad Xiao-Rui LYU$^{1,2}$
\quad Shi CHEN$^{1}$
\quad Wen-Qian HUANG$^{1}$
\quad Dao-Jin HONG$^{1}$\\
\quad Yang-Heng ZHENG$^{1,2}$
}
\maketitle

\address{%
$^1$ University of Chinese Academy of Sciences, Beijing, 100049, China\\
$^2$ CAS Key Laboratory of Vacuum Physics, UCAS, Beijing 100049, China\\
$^3$ Guangxi Key Laboratory for the Relativistic Astrophysics, Guangxi University, Nanning 530004, China\\
$^4$ Institute of High Energy Physics, Chinese Academy of Sciences, Beijing 100049, China
}

\begin{abstract}

A sealed high gas pressure detector working in pure argon is assembled. It consists of a 5 cm $\times$ 5 cm PCB THGEM (THick Gaseous Electron Multipliers). The detector structure and experimental setup are described.
The performances under high pressure of 2 atm mainly consist in selecting optimal voltages for ionization region and induction region.
The dependence of the shape of Alpha particle spectra measured with relative gas gain on gas pressure (1.3 $\sim$ 2.0 atm) has been studied.
The 8 groups of relative gas gain versus working voltage of THGEM expressed by weighting filed $E/P$ are normalized, being consistent with theory.
The results show that the air tightness of the chamber is good measured by a sensitive barometer and checked with gas gain.
The experimental results are compared with Monte Carlo simulation on energy deposition without gas gain involved.
\end{abstract}



\footnotetext[0]{\hspace*{-3mm}\raisebox{0.3ex}{$\scriptstyle\copyright$}2013
Chinese Physical Society and the Institute of High Energy Physics
of the Chinese Academy of Sciences and the Institute
of Modern Physics of the Chinese Academy of Sciences and IOP Publishing Ltd}%

\begin{multicols}{2}

\section{Introduction}

Gaseous detectors has been widely used in collider physics and other science frontier due to it's large volume with good spatial resolution and a reasonable cost.
Classically, most gaseous detectors should be flushed by continuous working gas. 
In some special fields, for instance, astrophysics or outdoor applications, the cumulative volume is limited while good performance of detector is required. A sealed gaseous detector will be one of the candidates to meet those requires.

The THick Gaseous Electron Multiplier (THGEM)~\cite{THGEM1,THGEM2,THGEM3} as an "expanded" GEM explored by A. Breskin et al in $2004$ is economically produced mainly on printed circuit board (PCB).
In recent years, some new substrates (ceramic and Liquid Crystal Polymer (LCP) etc.) and new technologies (laser hole-drilling and chemical etching) have been developed very rapidly.
THGEM has the excellent characteristics of robust, low cost, high gain ($10^3\sim10^5$), good spatial resolution (sub-mm), high counting rate, good stability and easier manufacturing.
A special type of thinner-THGEM~\cite{hb,thinnerTHGEM,qian} was domestically developed in $2011$.

For a sealed gaseous detector, one of the most important issues is to keep the parameters of the working gas as constant as possible.
Therefore, our first task is to study the long-time stability of inner working gas of the sealed chamber.
In general, the air tightness of the sealed detector and outgassing rate of the detector's materials are influencing factors of long-time stability.

To test the air tightness of the detector, a barometer is used to monitor the change of the detector's gas pressure. On the other hand, an Alpha source $^{241}$Am placed inside the detector is used both for checking gas pressure by measuring the gas gain information from the Alpha spectra and for further studying the dependence of gas pressure on Alpha spectra.

Spectra sizes are important for different particles, for instance, heavy particles, light particles as well as photons of different energies. 
For the next step of developing novel detectors, in particular for non-ageing gaseous detector for visible light, using pure argon is very crucial~\cite{Bressan,Buzulutskov}.
In addition, the detector working under high gas pressure is very beneficial and advantageous to improve the detection efficiency and spatial resolution and to catch more effective track lengths in the fiducial volume of the detector~\cite{korff,sauli}.

Response of the deposited energy distribution of Alpha particles with its spectrum influenced by source-chamber geometry and gas pressure should be known for any heavy particles and light particles required in many related experiments,  such as some authors focused on alpha particle in environmental ambient gas~\cite{Charpak}, on geometrical relationship between source and collective electrodes of chamber~\cite{nine}, on UV photons under low temperature~\cite{Paredes}, on alpha spectra in different rare gases~\cite{Saito}, on the deposited energy distribution within different Bragg's curve sections for alpha and to adjust optimal ionization thickness for heavy particles~\cite{Kim}. All these mentioned issues have  shown the significance in this detector study field.

This paper is focused on the parameters of the THGEM detector working in high pressured pure Ar, the related techniques for keeping air tightness, the shapes of Alpha spectra under different high pressures, the long term stability, and the gas gain influenced by gas pressure. Monte Carlo simulations with GEANT4 software~\cite{geant} and calculations are compared with the experimental results.

\section{Experimental setup}

As illustrated in Fig.~\ref{fig:physical}, a sealed gaseous detector based on THGEM has been built.
The key device of the detector is a single layer of 5 cm $\times$ 5 cm PCB THGEM board, $200$ $\mu$m thickness, claded with 18 $\mu$m of copper and plated with gold layers on the top of the Cu film. The holes are cylindrical (diameter $200$ $\mu$m) with $500$ $\mu$m pitch. They are produced at University of Chinese Academy of Sciences (UCAS) and the Second Academy of China Aerospace Science and Industry Corporation (CASIC).
The structure and characteristic of THGEM is seen in reference~\cite{qian}.
The distance between the cathode mesh and the upper layer of the THGEM, which is the drift region, is 1.5 cm, while the distance between the lower layer of the THGEM and the anodes, which is the induction region, is 0.3 cm.

\begin{center}
\includegraphics[width=6cm]{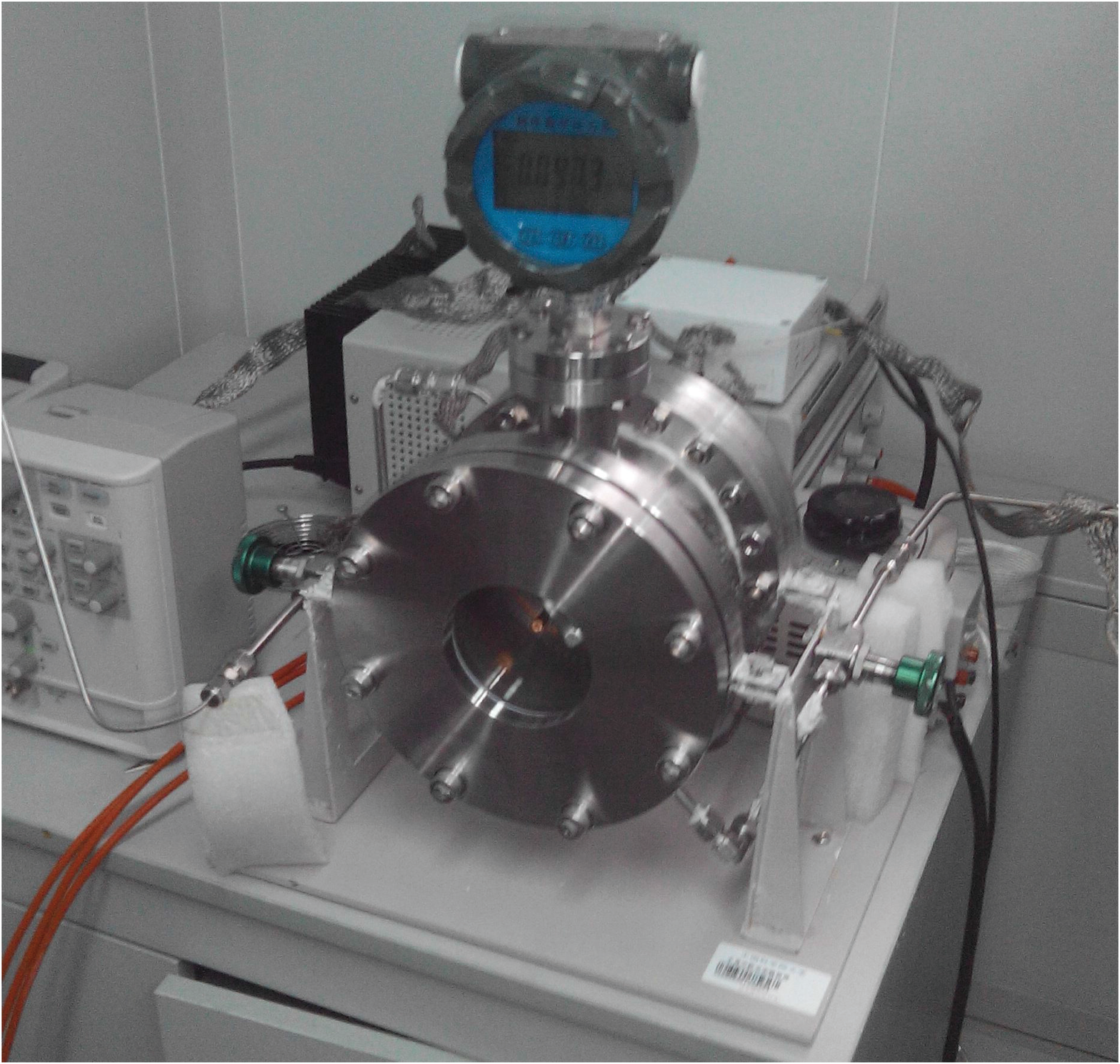}
\figcaption{\label{fig:physical}   The facility of the sealed detector. }
\end{center}

The shell of the detector is a stainless-steel cylindrical vessel. The chamber is sealed by vacuum flanges tightly connected with oxygen-free copper and rubber ring. There are 8 Kovar electrodes on the bottom flange, which are used for providing high voltage feedthrough. The anode's readout consists of the vacuum multi-pin plug-and-socket being able to meet the future purpose, and temporarily being connected as one output. The front end window is comprised of an ordinary glass of $\varnothing8\times0.8$ cm. The glass is tightly pressed with stainless-steel chamber by Teflon.
With the glass window, the detector is able to detect outer source, especially designed for visible light.
Two precise aciculiform valves are used on both sides of the flanges to guarantee the air tightness. The chamber leakage has been checked with He gas probe and deaerated under 100 $^{\circ}$C after being assembled. The result shows that the vacuum degree of the chamber could attain down to $10^{-5}$ Pa, and the leak rate of chamber is less than 10$^{-10}$ Pa$\times$m$^{3}$/s. The various parts of the sealed detector are easy to be assembled shown in Fig.~\ref{fig:chamber}.

\begin{center}
\includegraphics[width=6cm]{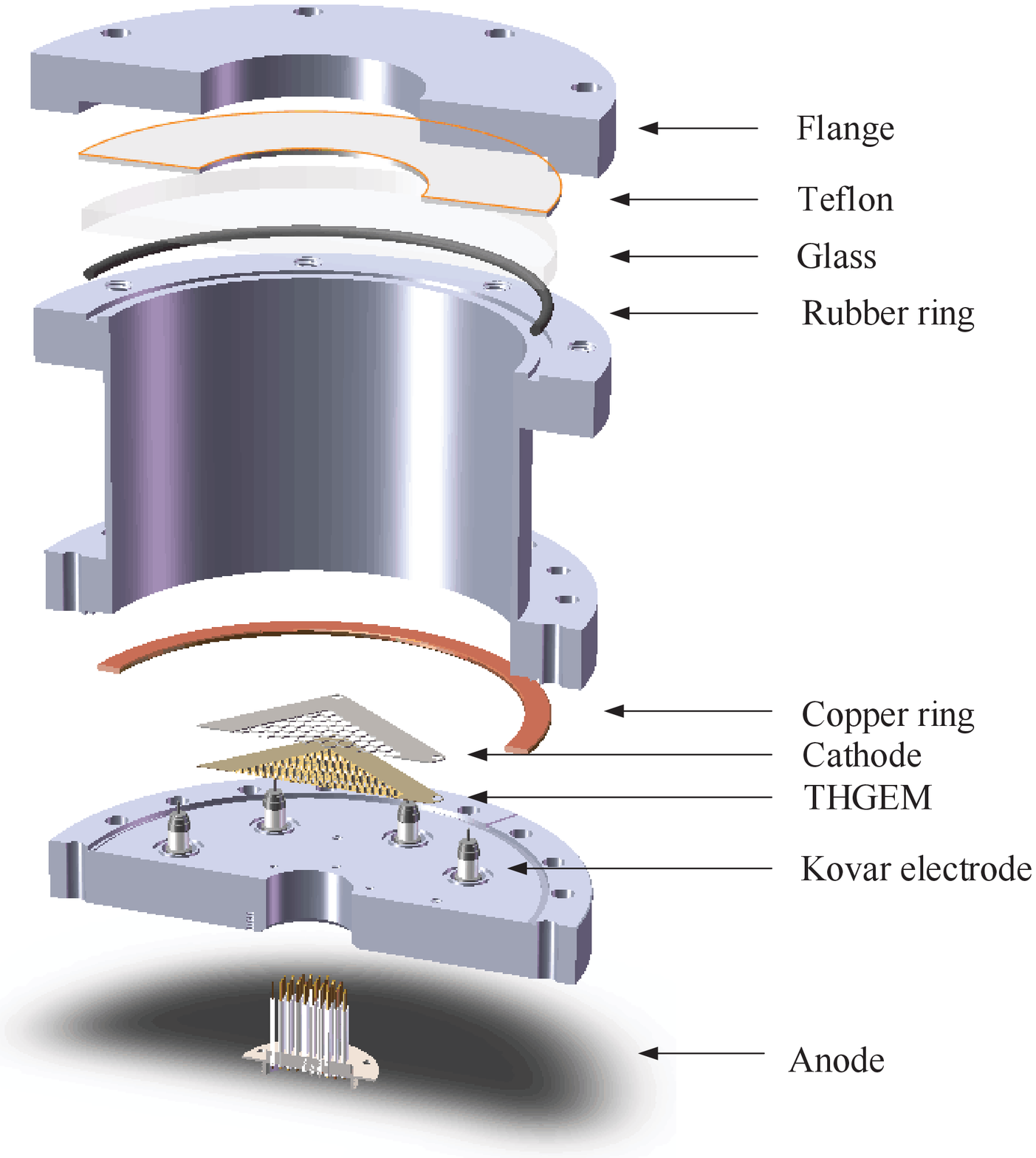}
\figcaption{\label{fig:chamber}   The structure of the sealed chamber. }
\end{center}

The high voltages on the electrodes are provided by a CAEN NDT1471H HV programmable power supply which  provides 4 independent High Voltage channels. The signals are recorded by a CR110 pre-amplifier (charge sensitivity 1.4 V/pC) followed by a CAEN N968 spectroscopy amplilier (shaping time is 0.5 $\mu$s, amplified factor is 100) and a CAEN N957 8K Multi-Channel Analyzer (MCA, range $0 \sim 10$ V). A schematic diagram of the structure of the detector the test setup is shown in Fig.~\ref{fig:Setup}.

\begin{center}
\includegraphics[width=8cm]{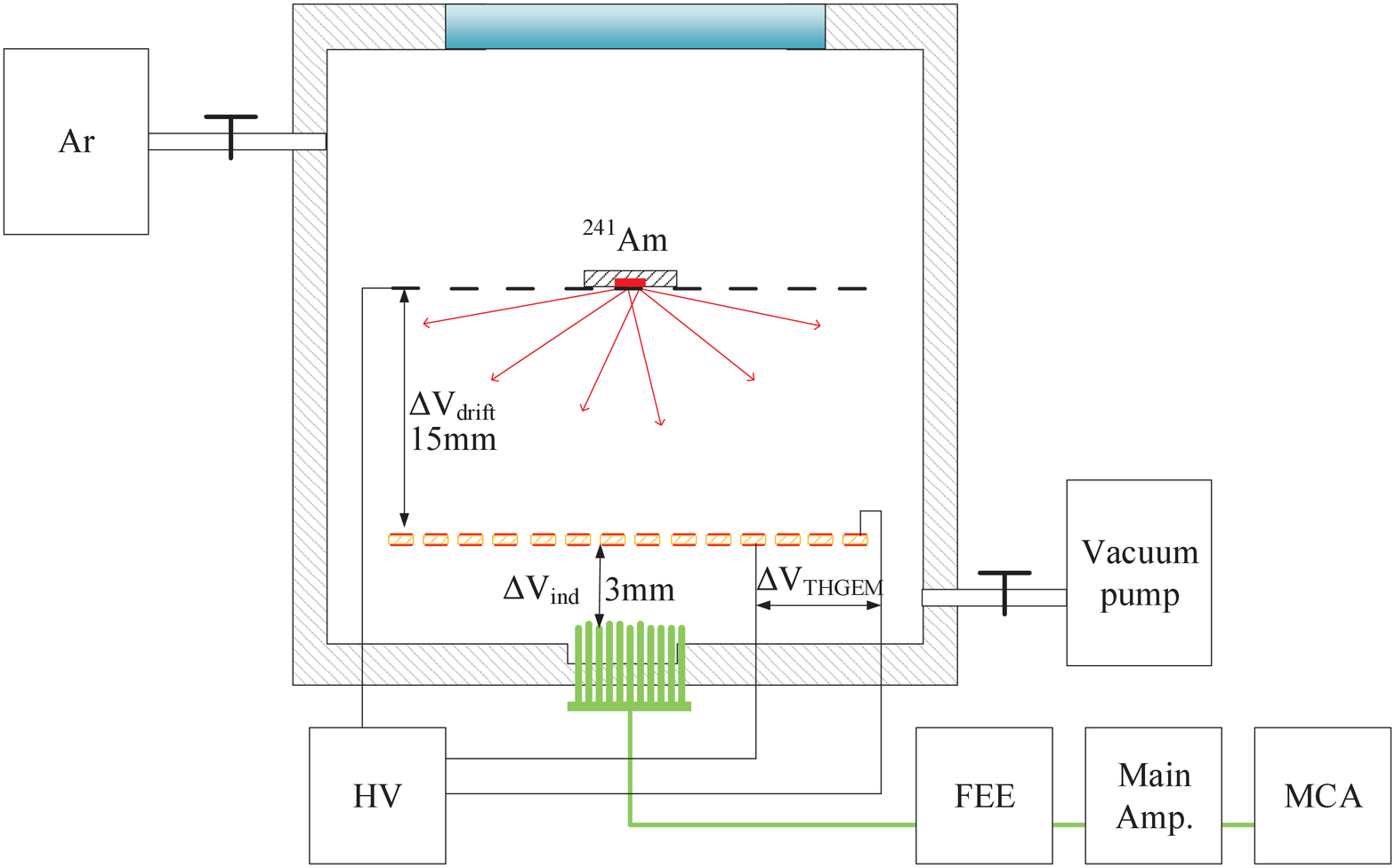}
\figcaption{\label{fig:Setup}   Schematic diagram of the experimental setup. }
\end{center}

To test the long-time stability of the detector, an alpha source $^{241}$Am of 0.8 $\mu$Ci fixed at the centre of the cathode with an iron wire could be seen in Fig.~\ref{fig:Setup}. For $^{241}$Am, the half-life is about 433 years. So the activity is almost constant during the experiment. Those particles emit into the drift region with a solid angle of $2\pi$. During the whole period of the experiment, the spectra of Alpha particles are measured per day. The relative value of chamber's gas pressure compared to the ambient atmosphere is measured by a barometer connected to the chamber via a CF35 flange. The resolution of the barometer is 2.5 kPa. The gain of the detector also could be affected by the environmental factors such as temperature and humidity~\cite{Amaro,Dong,zhou}, which are monitored by a sensor.

\section{Experimental results and discussion}

\subsection{Performance study under high pressure}

In order to guarantee the safety, especially to avoid the glass window from being broken, to test the anti-press strength of the detector chamber is the first step. The chamber is vacuumized till $10^{-9}$ atm (1 atm = $1.014\times10^5$ Pa) by molecular pump, and filled with pure Ar till 2.0 atm.
Then, the voltages are acted successively on all electrodes.

The performances under high pressure mainly consist in selecting optimal voltages in ionization region $\Delta V_{drift}$ and induction region $\Delta V_{ind}$.

At the beginning, a typical MCA spectra of Alpha particles is measured under 2.0 atm shown in Fig.~\ref{fig:700alpha} with $\Delta V_{drift} = 400$ V, $\Delta V_{THGEM} = 700$ V and $\Delta V_{ind} = 600$ V. Except for the low range electronics noise, a pulse peak is clearly seen.
The gas gain of the detector could be represented by the channel number of the position of the pulse peak.

\begin{center}
\includegraphics[width=8cm]{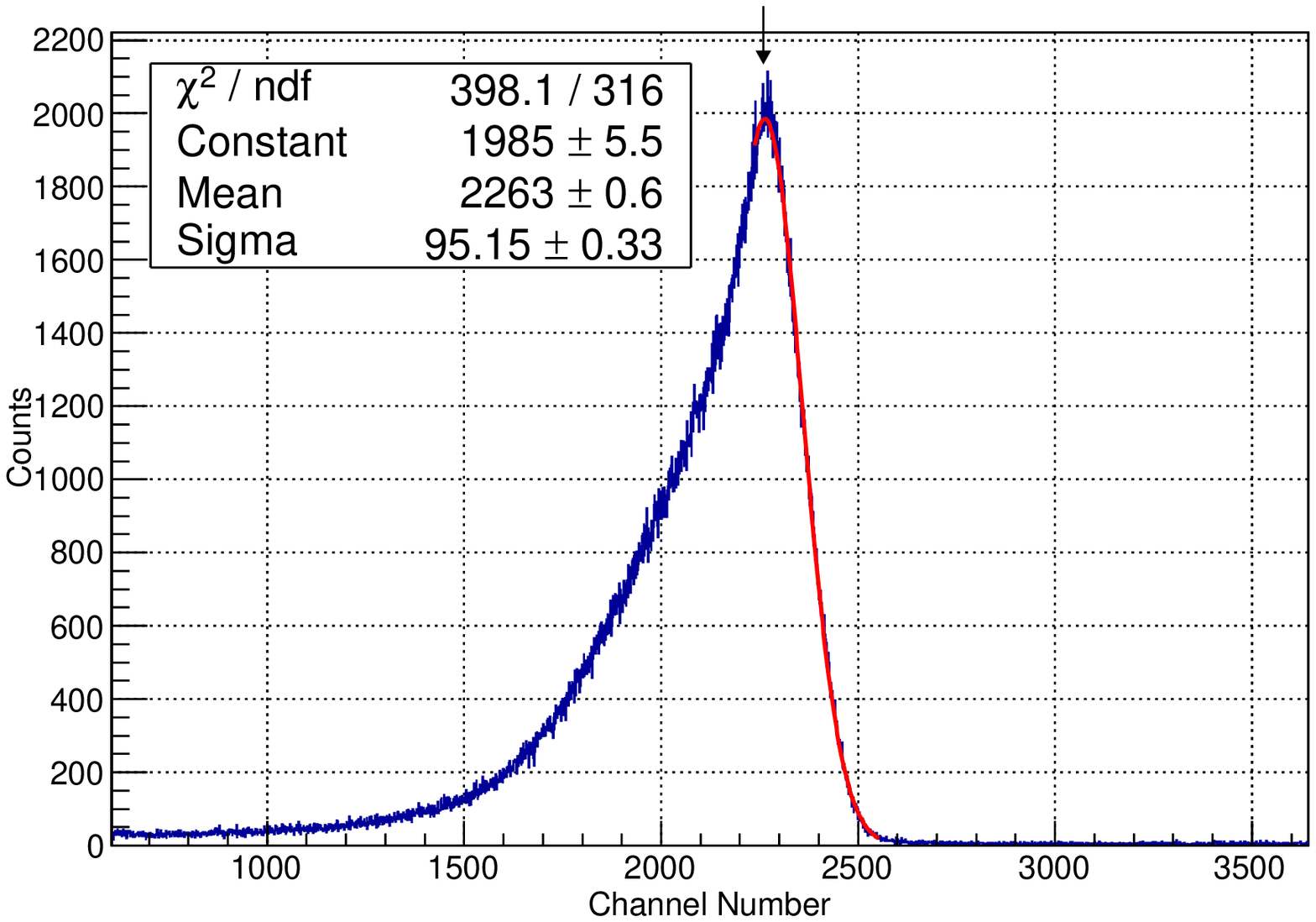}
\figcaption{\label{fig:700alpha}   A typical MCA spectrum measured by the chamber's detector at 2.0 atm. }
\end{center}

From the relative gains (channel number) versus the drift voltage ($\Delta V_{drift}$) under the same working voltage ($\Delta V_{THGEM} = 700$ V) and the same induction voltage ($\Delta V_{ind} = 600$ V) shown in Fig.~\ref{fig:cathode2}, the optimum $\Delta V_{drift}$ is 400 V with respect to the maximum gas gain.
For Fig.~\ref{fig:cathode2}, the curve is caused by two factors.
With increasing $\Delta V_{drift}$, the recombination effect reduces, and the more primary ionized electrons appear.
The other factor is that with increasing $\Delta V_{drift}$, the efficiency of electron entering holes on the upper layer of the THGEM reduces.

\begin{center}
\includegraphics[width=8cm]{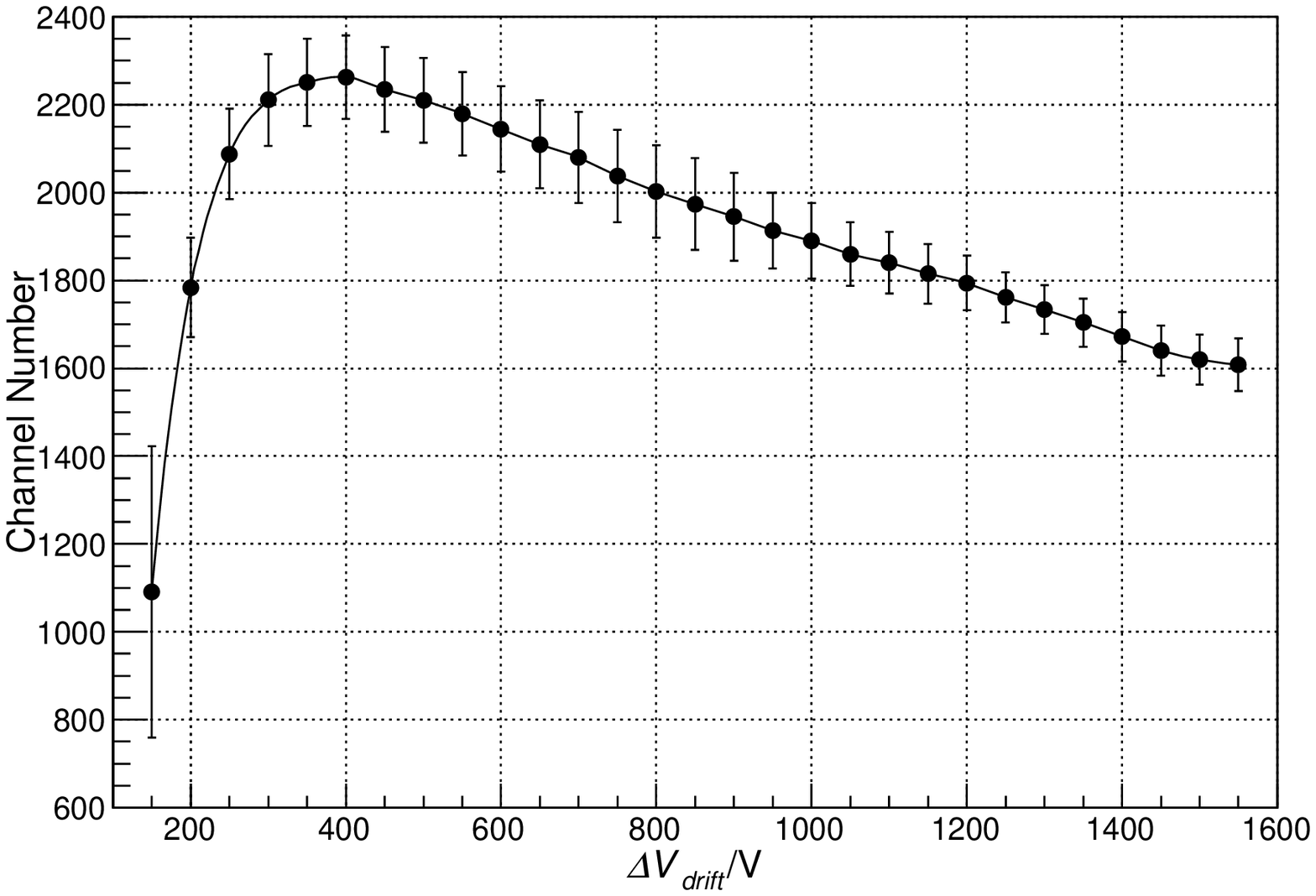}
\figcaption{\label{fig:cathode2}    The channel number of the pulse peak positions changes with $\Delta V_{drift}$ under fixed $\Delta V_{THGEM}$ and $\Delta V_{ind}$. }
\end{center}

As illustrated in Fig.~\ref{fig:thgemdown1}, the relative gains of the different $\Delta V_{ind}$ under $\Delta V_{THGEM} = 600$ V and $\Delta V_{drift} = 400$ V are measured. As the special structure of the anode, the optimal value of $\Delta V_{ind}$ exists.
The gas gain of THGEM is also positive correlated with electron collection efficiency~\cite{Ropelewski,Alexeev,wang}. Due to the effect of ion back flow, the electron collection efficiency would not always improve with increasing $\Delta V_{ind}$.

\begin{center}
\includegraphics[width=8cm]{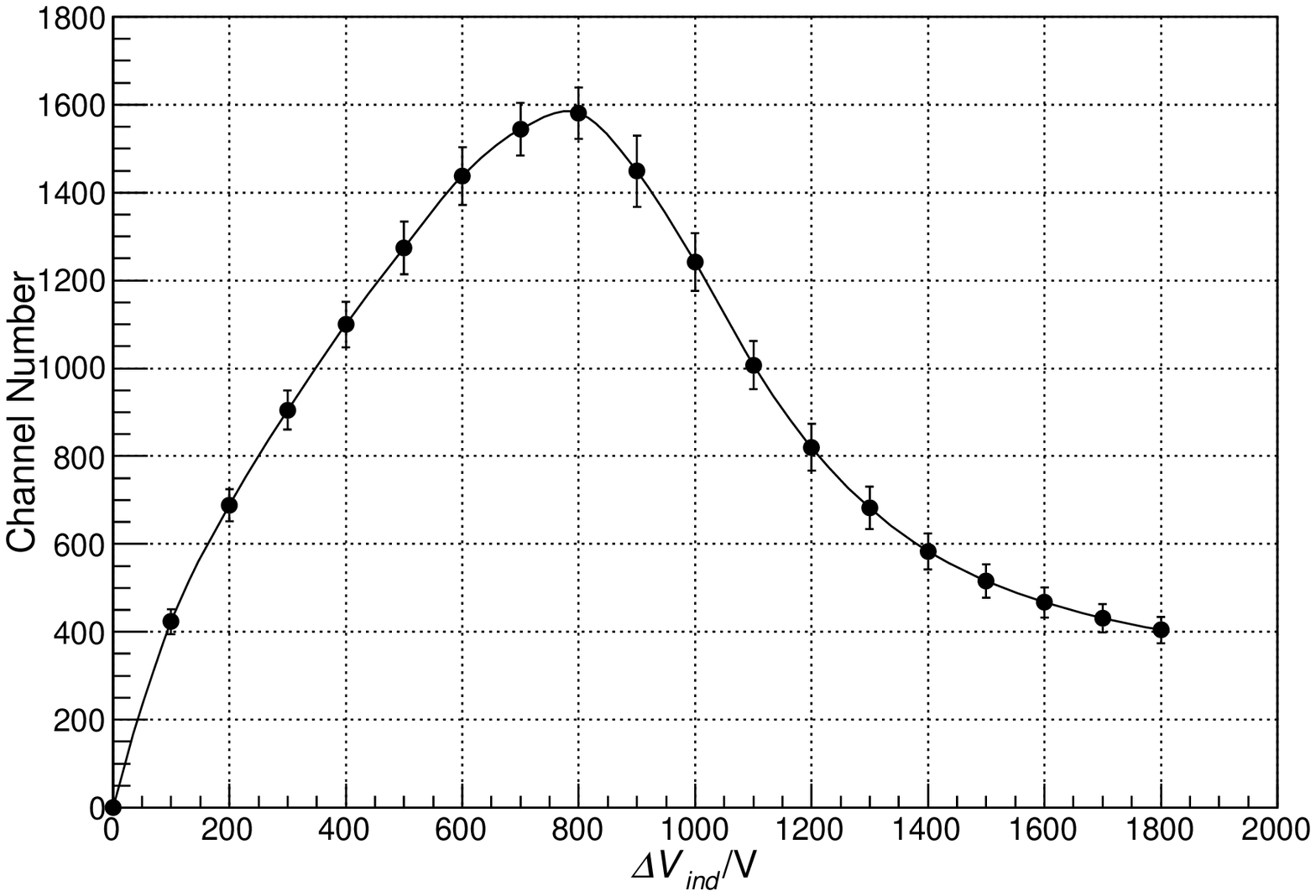}
\figcaption{\label{fig:thgemdown1}   The channel number of the pulse peak changes with $\Delta V_{ind}$ under fixed $\Delta V_{THGEM}$ and $\Delta V_{drift}$. }
\end{center}

\subsection{Alpha particle spectra under various gas pressure}

The $^{241}$Am source emits Alpha particles with energies of $\sim$ 5.486 MeV ($\sim$ 85.2\%) and $\sim$ 5.443 MeV ($\sim$ 12.8\%)~\cite{alpha}.
For the 5.486 MeV Alpha particles, their range in ordinary air is $\sim$ 3.5 cm.
According to Bragg-Kleeman rule, the estimated range of the Alpha particle in the normal temperature-pressure Ar is $\sim$ 4.26 cm.
For the $^{241}$Am source located at the top centre of the cathode surface, only parts of deposited energies of the Alpha particles are catched in the fiducial volume.
It means, if Alpha particles are required completely ionizing in the drift region, the pressure of the detector should be increased. To verify this deduction, the MCA spectra of Alpha particles for different values of gas pressure are obtained.

A series of Alpha-ray spectra of 8 groups of various pressures (1.3 $\sim$ 2.0 atm) under 5 different $\Delta V_{THGEM}$ have been measured within 6 hours of stable environment conditions (see in appendix).

Two MCA spectra of Alpha particles measured at 1.3 atm and 2.0 atm are shown in Fig.~\ref{fig:low} and Fig.~\ref{fig:high}, in which, three group of voltages are the same ($\Delta V_{drift} = 400$ V, $\Delta V_{THGEM} = 600$ V and $\Delta V_{ind} = 600$ V). In Fig.~\ref{fig:pressure},  an obvious tendency shows that, under the same $\Delta V_{THGEM}$, the higher the pressure of the chamber ($1.8 \sim 2.0$ atm), the smaller the gas gain.

Fig.~\ref{fig:low} and Fig.~\ref{fig:high} are explained as the following:
\begin{enumerate}
  \item With decreasing pressure, part of Alpha particles escape out of the fiducial (sensitive) volume from lateral sides, and double peaks appear, just caused by these shorter remained tracks (Fig.~\ref{fig:low}).
  \item While the pressure is very high, all particles stop (including Bragg's tail) in the fiducial volume, and single peak appears (Fig.~\ref{fig:high}).
  \item From 8 groups of various pressure (1.3 $\sim$ 2.0 atm), the evolutional tendency of the Alpha spectra sizes is obvious: both peak (max channel) and starting point (min channel) with a valley between two peaks decrease, i.e. shift to lower channels with increasing pressure. Fig.~\ref{fig:pressure} summarizes the channel number changes with $\Delta V_{THGEM}$ under 3 groups of different pressures.
      This results could be explained by 2 factors: firstly, even more energy deposited in the fiducial volume with pressure increasing ($dE/dx$ proportional to gas density); secondly, the gas gain is a function of weighting electric field strength $E/p$, being smaller in the THGEM holes. This effect causes decrease of gas gain which would be the dominant factor compared to the first.
\end{enumerate}

\begin{center}
\includegraphics[width=8cm]{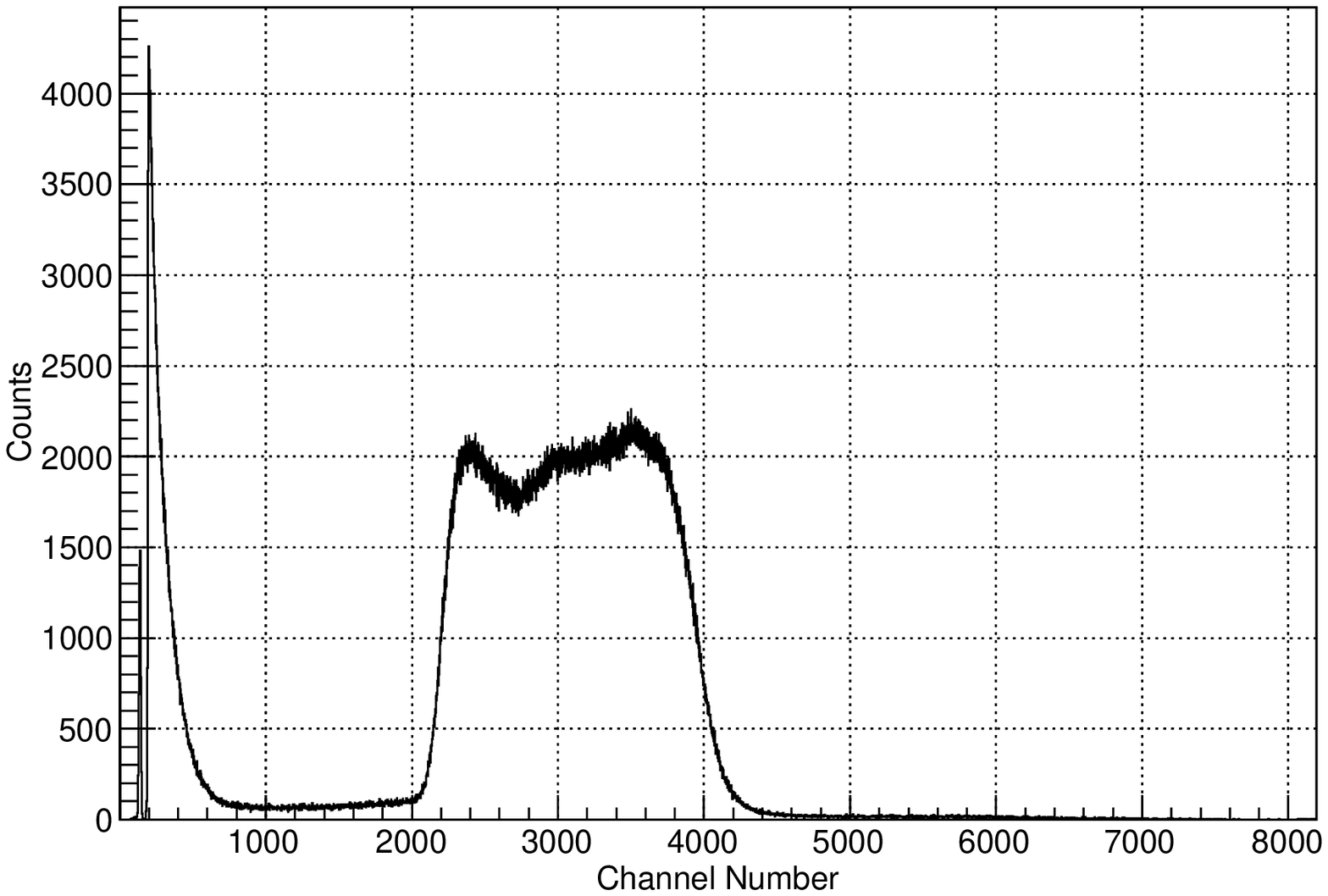}
\figcaption{\label{fig:low}   The MCA spectrum of Alpha particles measured at 1.3 atm. }
\end{center}

\begin{center}
\includegraphics[width=8cm]{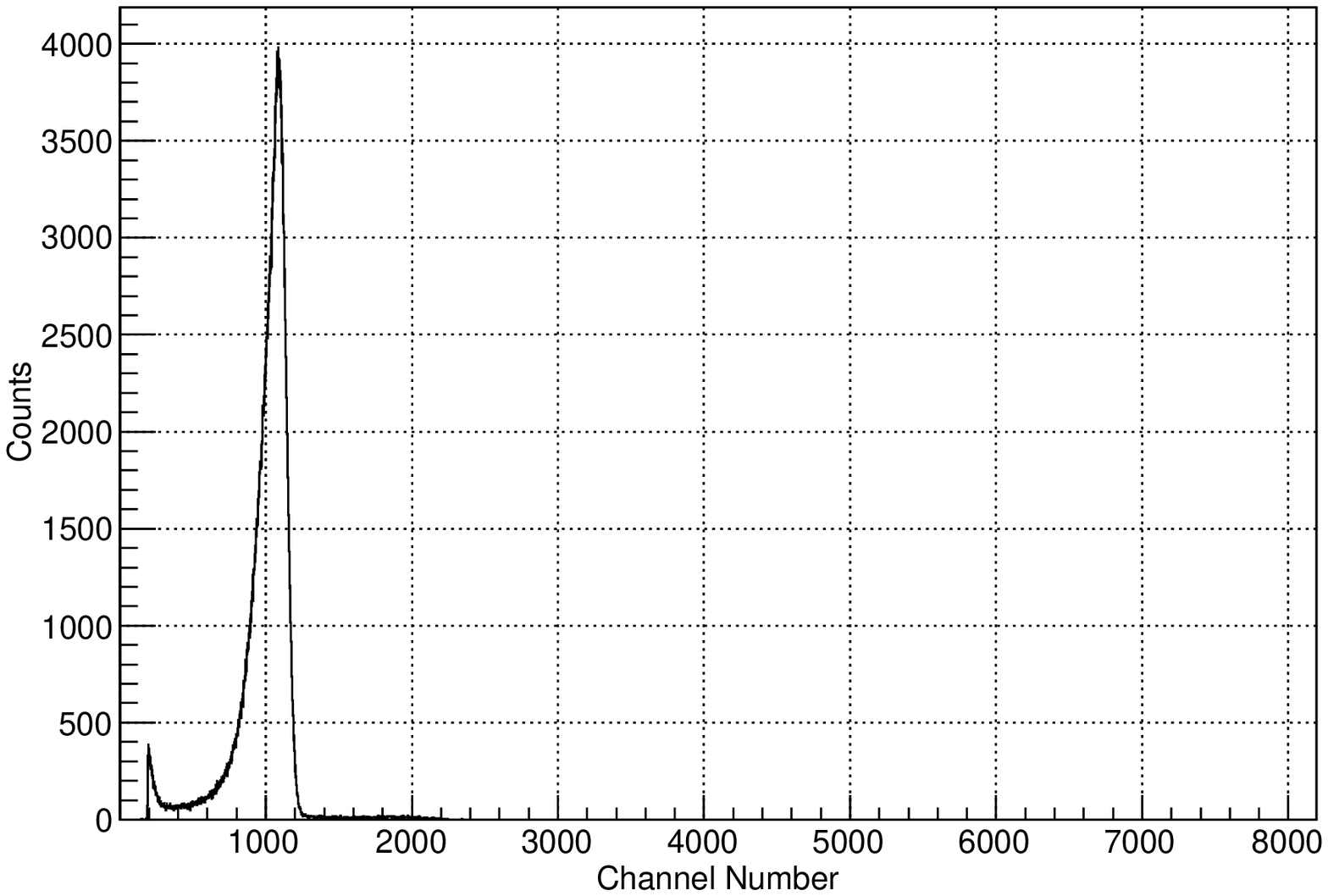}
\figcaption{\label{fig:high}   The MCA spectrum of Alpha particles measured at 2.0 atm. }
\end{center}

\begin{center}
\includegraphics[width=8cm]{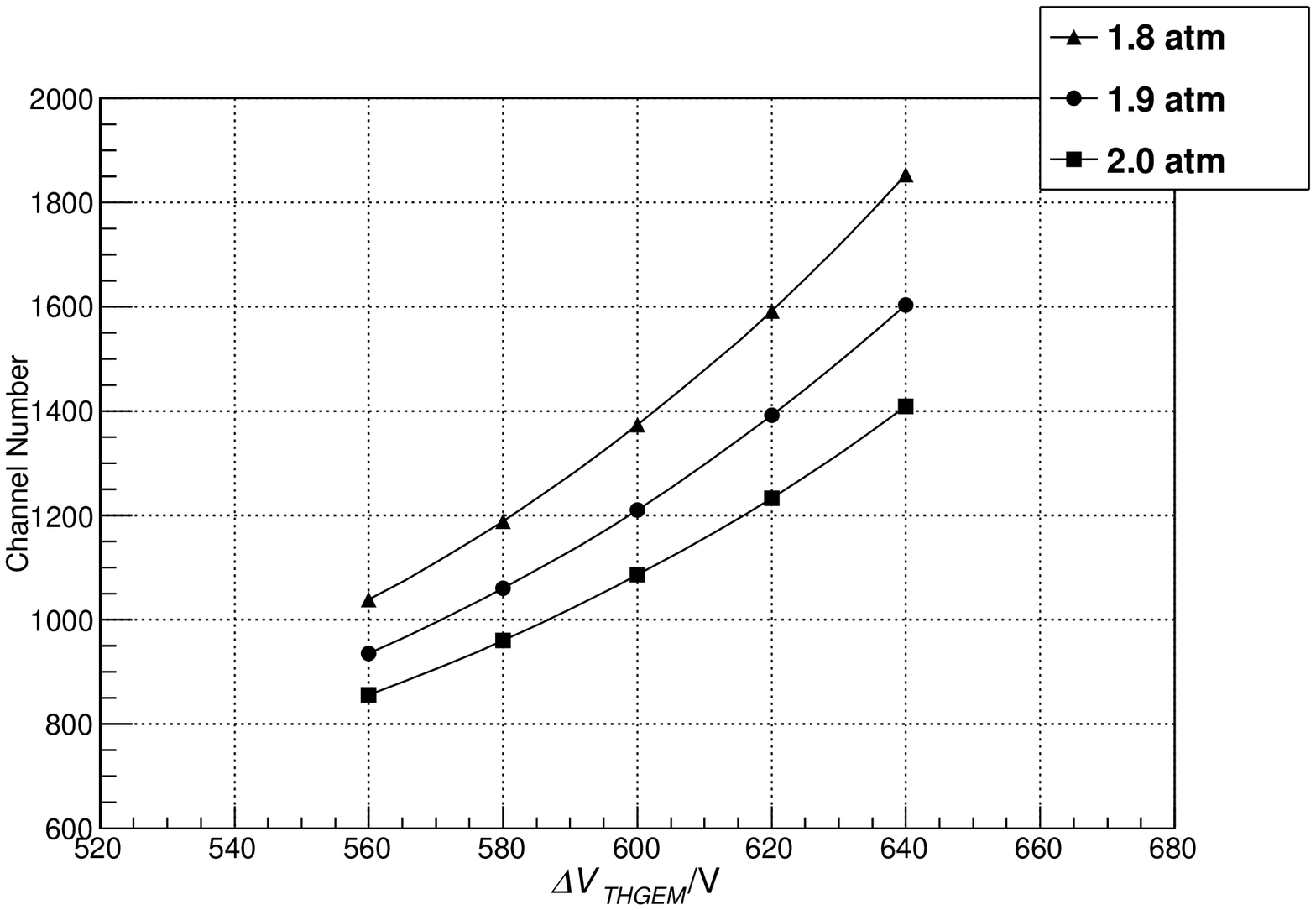}
\figcaption{\label{fig:pressure}   At different inner pressure of the chamber, the channel number of the pulse peak changes with $\Delta V_{THGEM}$. }
\end{center}

From the gas amplification theory developed by Townsend~\cite{sitar}, the gas amplification coefficient, or gas gain $M$ in high electric field is related to the first Townsend ionization coefficient $\alpha_T$. $1/\alpha_T$ is the mean free path between ionizing collisions. In quite rough approximation, the avalanche multiplication of electrons in THGEM could be deduced by using the parallel-plate approach as $\ln M/(Pd)$, where $d$ is the interelectrode distance, $P$ is the pressure, i. e. the weighting electric field of the THGEM is $E/P \approx \Delta V_{THGEM}/(Pd)$, seen in reference~\cite{Buzulutskov}. For 40 spectra in 8 groups of different pressures, the gas gain could be represented by the channel number $N_{Ch}$ of the pulse peak (to the spectra of two peaks, the right one is selected).
Values of $\ln N_{Ch}/(Pd)$ as a function of the weighting electric field $V_{THGEM}/(Pd)$ are shown in Fig.~\ref{fig:townsend}. The tendency of $\ln N_{Ch}/(Pd)$ conressponding to gas amplification changing with the weighting electric field is expressed as normalized linearity.

\begin{center}
\includegraphics[width=8cm]{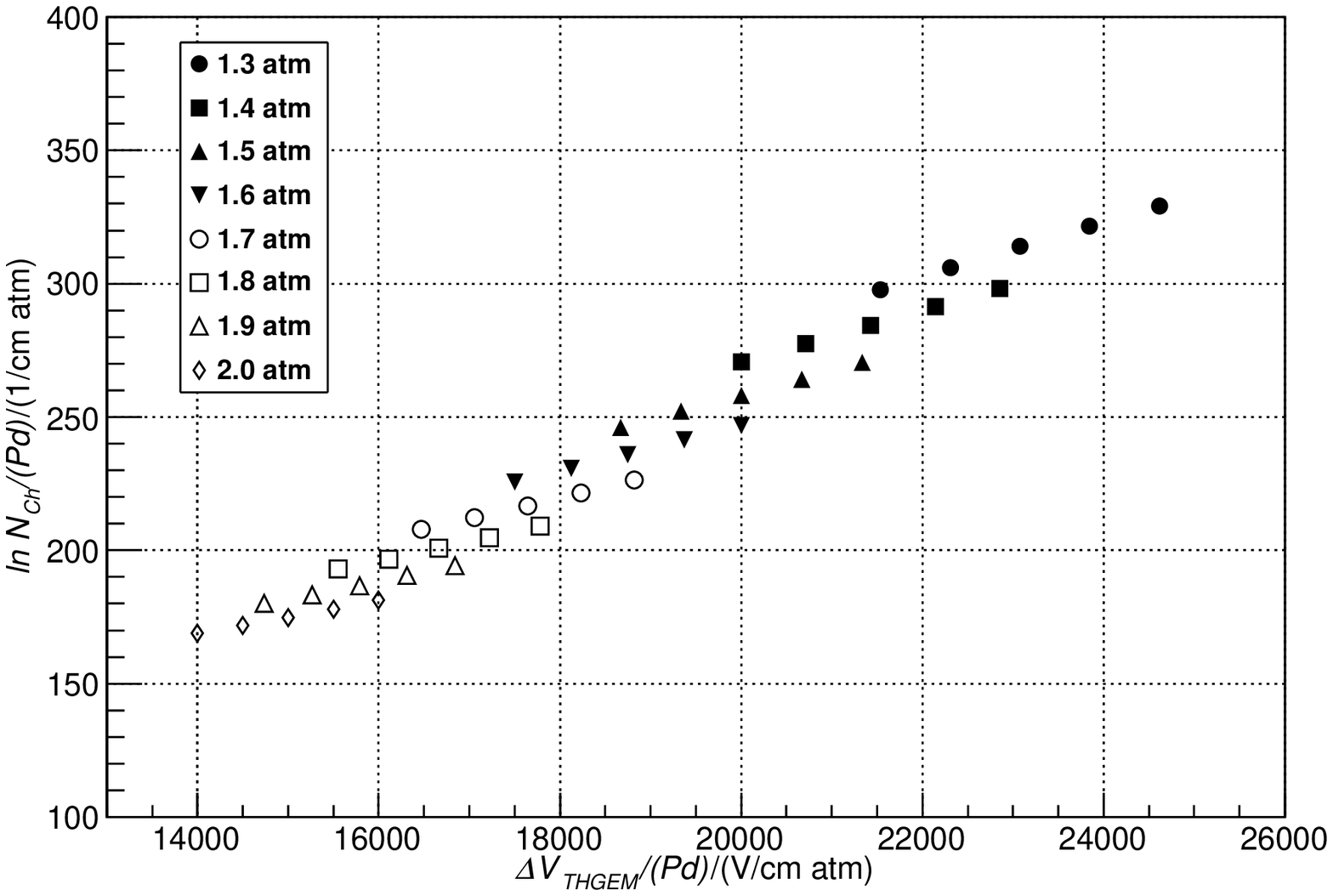}
\figcaption{\label{fig:townsend}   Formulae for $\Delta V_{THGEM}/(Pd)$ and for the relative gas amplification expressed in the form $\ln N_{Ch}/(Pd)$. }
\end{center}

\subsection{Study of long term stability and gas gain response}

To study the long-time stability of gain, we repeatedly obtained the spectra of Alpha particles measured by the detector working at 1.5 atm during a long term period. Under the conditions of $\Delta V_{THGEM} = 600$ V, $\Delta V_{drift} = 1000$ V and $\Delta V_{ind} = 600$ V, the MCA spectrum of Alpha particles has a pulse peak like Fig.~\ref{fig:700alpha}.
Considering that the temperature or other environmental factors might affect the gas gain, the experiment is selected during a period of the days with stable environmental conditions.
Under these conditions, the gain rose by $\sim5\%$ during $7$ days shown in Figure.~\ref{fig:day}.

\begin{center}
\includegraphics[width=8cm]{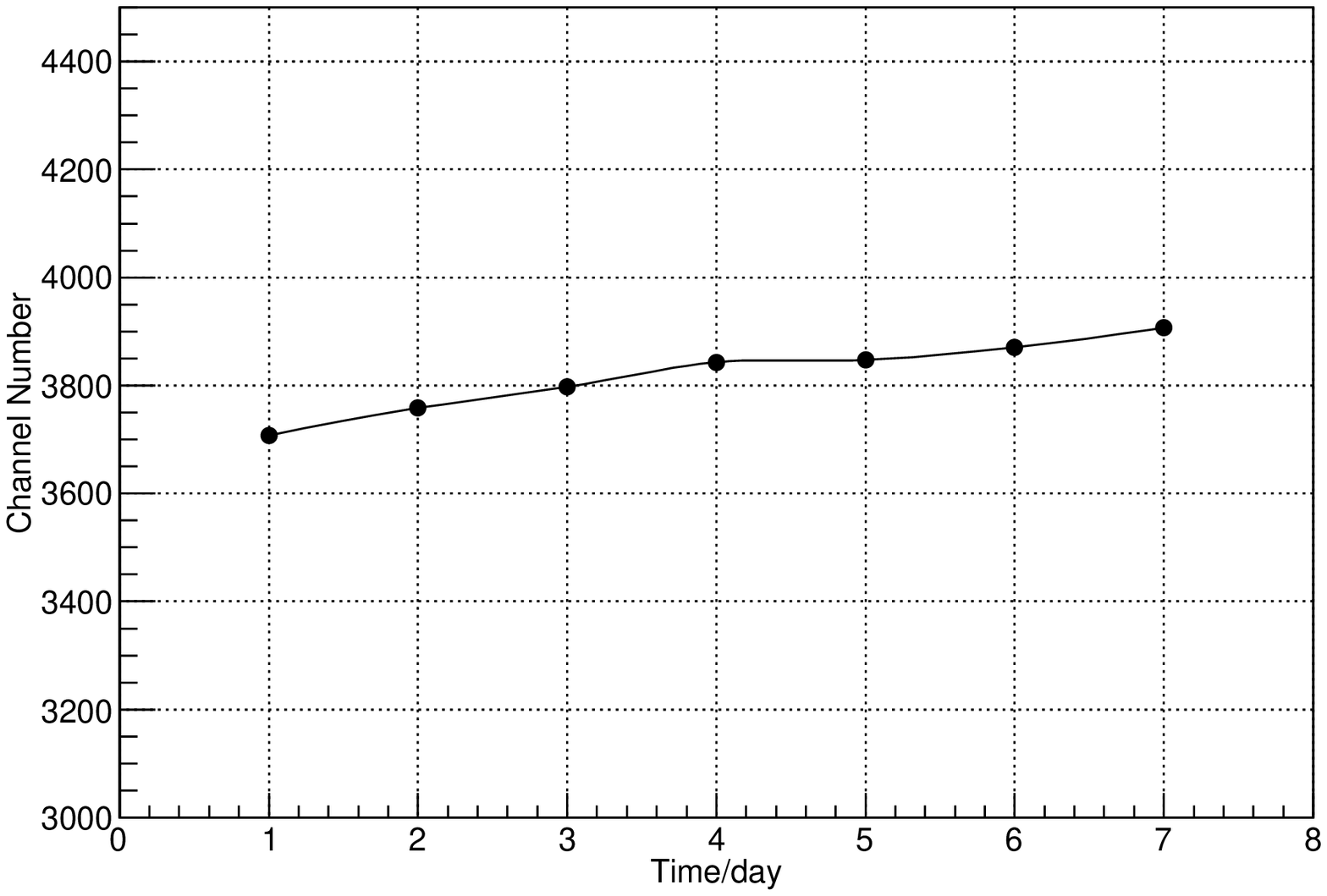}
\figcaption{\label{fig:day}   The channel number of the pulse peak measured by the detector at 1.5 atm changes with time, under the conditions of $\Delta V_{THGEM} = 600$ V, $\Delta V_{drift} = 1000$ V ang $\Delta V_{ind} = 600$ V. }
\end{center}

With the barometer (specification is 2.5 kPa) and the temperature sensor (specification is 0.5 $^{\circ}$C), the definite gas density of the sealed chamber is obtained during this period.
The change of the $p/T$ within 18 days is not beyond the measurement error 8.4 kPa/K (K is Kelvin scale). It means that the leakage of the chamber is very low.

In Fig.~\ref{fig:pressure}, the pressure changes from 2.0 atm to 1.9 atm, the gas gain rises $\sim11.4\%$. Roughly, if pressure changes 0.01 atm, the gain would changes 1\%.
Concerning the gas gain changing within relative long-term (7 days), the gas gain slightly increases with time shown in Figure.~\ref{fig:day}.
This tendency could be explained by some factors. One of factors is small leakage influencing the weighting field in the chamber.
Therefore, if the change of the gas gain could be controlled within $\sim10\%$, the stable period of the detector would be about 14 days.
In addition, the further long-term stability should be considered with the discharge and the outgas of the inner material of the chamber.

\section{Monte Carlo simulation and calculations}

To compare the different tendencies between experimental Alpha spectra size obtained by gas gain, and of
the energy deposition spectra obtained by GEANT4 simulation with the 5.486 MeV Alpha particles deposited in the fiducial volume of pure Ar at 3 different pressures, as shown as Fig.~\ref{fig:atm}. The emitting particles are set coming from a point source at the centre of the upper surface of the fiducial volume with a solid angle of 2$\pi$.

As Fig.~\ref{fig:atm} shows that the energies of the Alpha particles almost deposit in the pure Ar under 2.0 atm.
However, for the lower gas pressure, for instance, 1.0 atm or 1.5 atm, considerable numbers of Alpha particles do not loss all of their energies. For 1.0 atm, there is even no peak at 5.486 MeV, but two peaks at lower energies.
Obviously, for almost all Alpha particles, their energy deposited associated with the tracks in the fiducial volume could only cover parts of their ranges.
As the pressure increases, more and more tracks involve the whole ranges of the Alpha particles.
It will form the peak at 5.486 MeV in the energy spectrum.
Accordingly, the counts of the tracks in the lower energy region with two peaks gradually reduce and the peaks shift to high energy region, with a small part remained shown as the green histogram of 2.0 atm in Fig.~\ref{fig:atm}.
All these results are consistent with those obtained from experiments only without gas gain influence.

\begin{center}
\includegraphics[width=8cm]{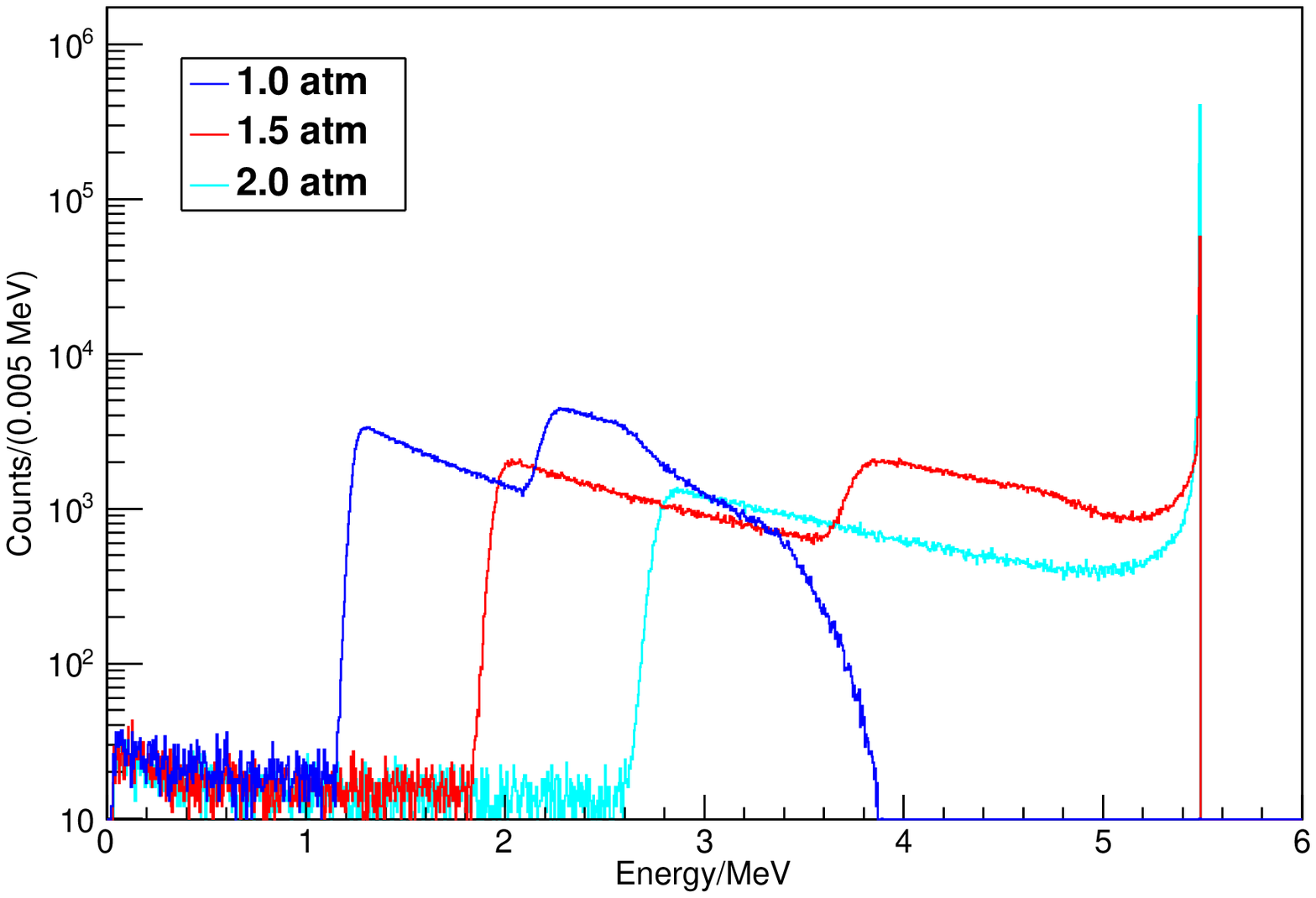}
\figcaption{\label{fig:atm}   The simulated energy spectra of the Alpha particles deposited in the fiducial volume at 3 different pressures. }
\end{center}

In case of considering the tracks of the Alpha particles which only point to the circular disc, and ignoring the others pointing to the lateral side or the corner of the drift region, the energy spectrum estimated by GEANT4 simulation as the red histogram shown in Fig.~\ref{fig:disc}.
The start point of the energy spectrum shown as the peak is due to the effect of shortest tracks, and the end part is contributed by the longest tracks ended at the edge of the circular disc.
The contribution of the tracks which point to the lateral side and the corner of the drift region stands for the green histogram of Fig.~\ref{fig:disc}.
An extra bump distribution appears at the largest MCA channel region which is contributed by those longest tracks ended at the corners of the fiducial volume with rectangular size.
By combining these two histograms to form the blue histogram of Fig.~\ref{fig:disc}, all the tracks are including in the fiducial volume.
The blue histogram has been obtained independently with all tracks in the whole fiducial volume.

\begin{center}
\includegraphics[width=8cm]{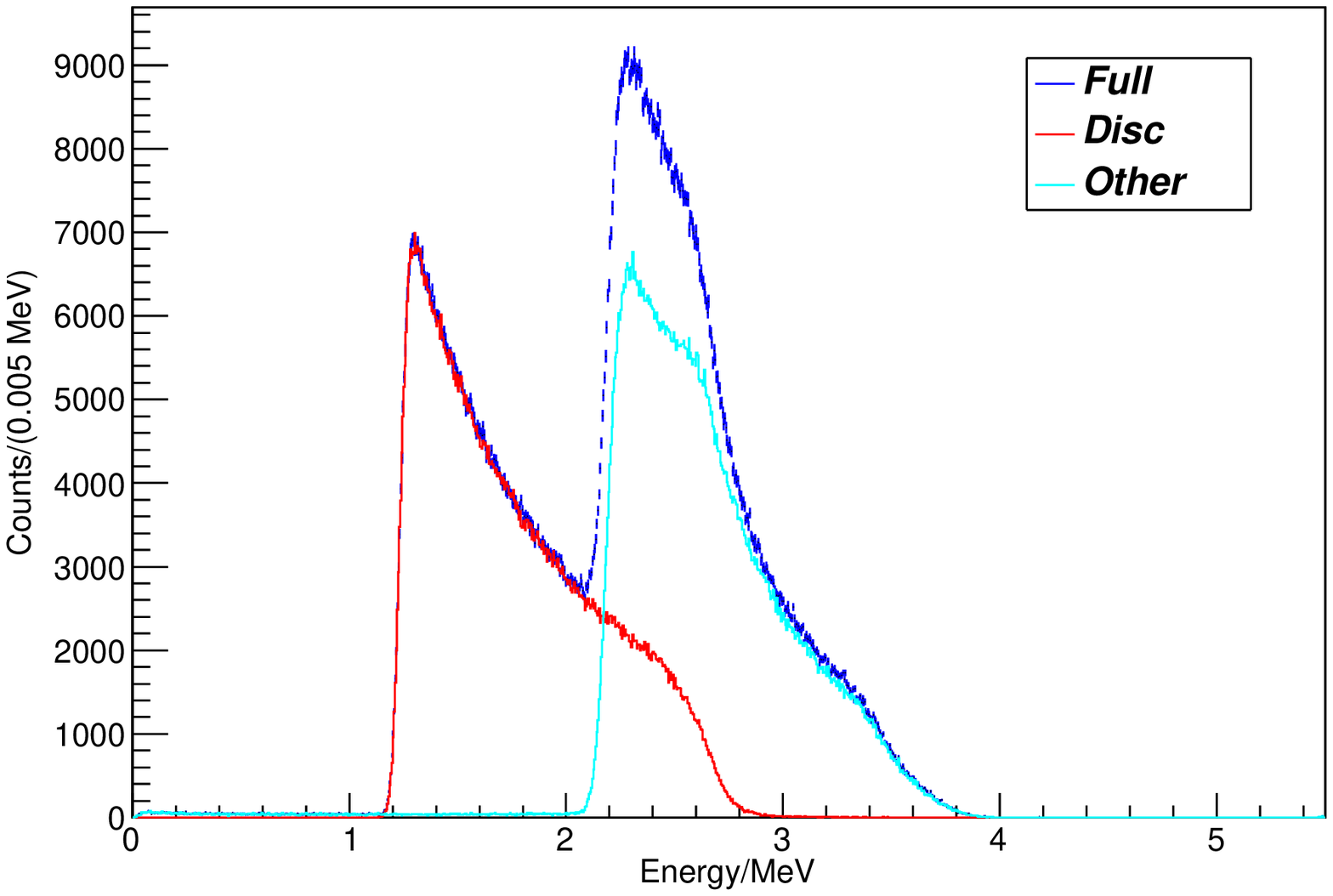}
\figcaption{\label{fig:disc}   All tracks energy distribution in the fiducial volume at 1.0 atm pressure estimated by GEANT4 simulation. }
\end{center}

Being different from the complicated rectangle fiducial volume, only a simple circular disc is considered, the solid angle $\Omega$ of a point source from the top center to the circular disc could be given by the following formula:
\begin{equation}
\Omega = 2\pi(1-\frac{h}{\sqrt{R^{2}+h^{2}}})
\end{equation}
where $R$ is the radius of the circular disc, and $h$ is height of the point source to the circular disc. The length of the point source to the edge of circular disc is $l = \sqrt{R^{2}+h^{2}}$, thus the differential is
\begin{equation}
\frac{d\Omega}{dl}=\frac{d\Omega}{dR}\frac{dR}{dl}=2\pi\cdot\frac{Rh}{(R^2+h^2)^{3/2}}\frac{l}{\sqrt{l^2-h^2}}=2\pi\cdot\frac{h}{l^2}
\end{equation}
where $R \leqslant 2.5$ cm and $h = 1.5$ cm. The differential changes with $l$, as Fig.~\ref{fig:domega} showing.
Form Fig.~\ref{fig:domega}, a large peak at the start point appears, then the curve gradually descends with increasing $l$, finally cuts off at the circular edge.

\begin{center}
\includegraphics[width=8cm]{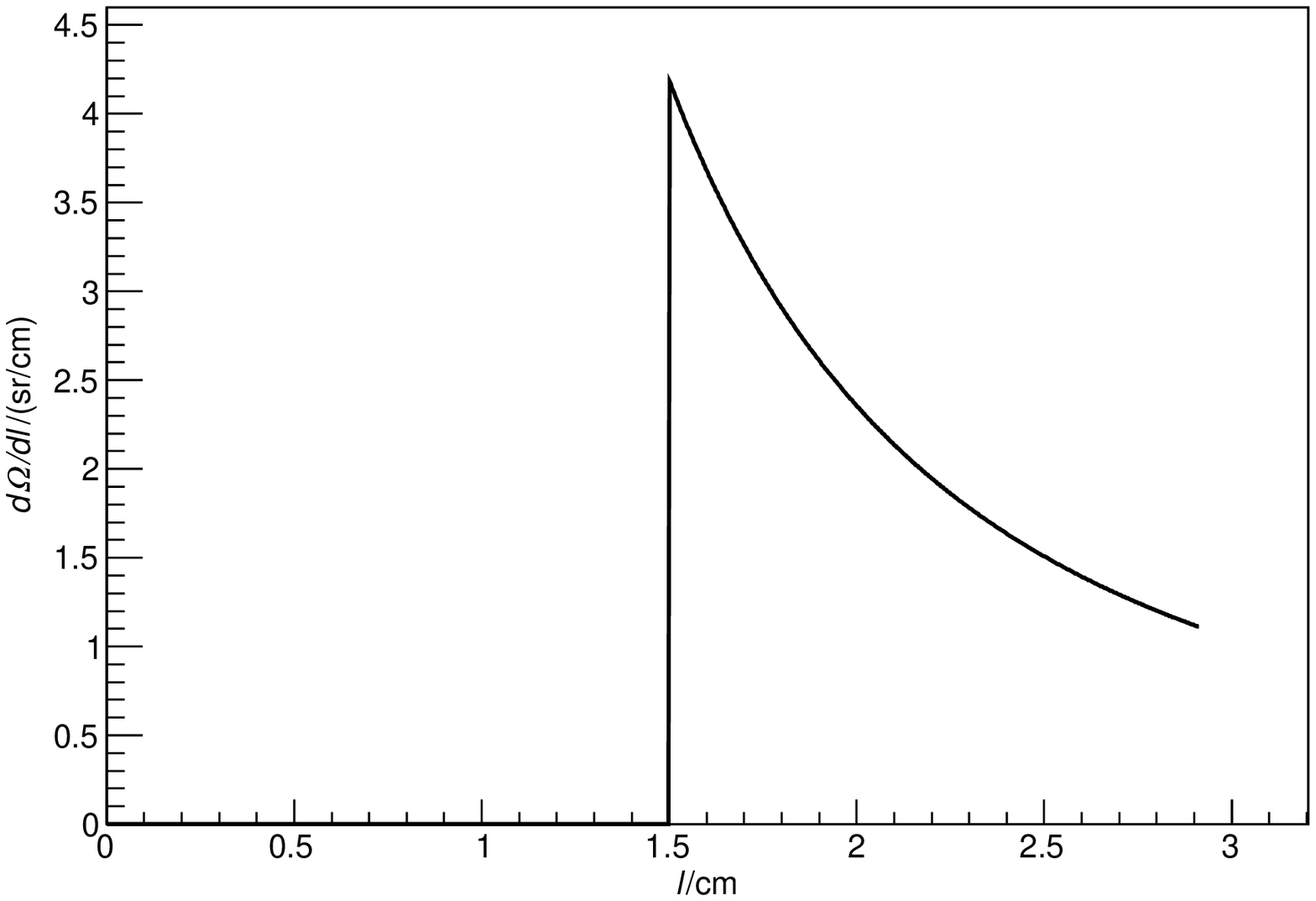}
\figcaption{\label{fig:domega}   $R \leqslant 2.5$ cm and $h = 1.5$ cm, the differential changes with $l$. }
\end{center}

$dN/dl=(dN/d\Omega)\cdot (d\Omega/dl)$, where $N$ is the number of the source, $dN/d\Omega$ is homogeneuous distribution ($dN/d\Omega=N_0/2\pi$, where $N_0$ is the initial value of $N$). Finally, this $dN/dl$ just represents the Alpha spectra, with constant $dE/dx$ along $l$. It coincides with Fig.~\ref{fig:disc}. After some factors including Bragg curve and non-point source size with film coating have been revised, this $dN/dl$ would be more precisely consistent with the experimental results.

\section{Conclusion and prospect}

In summary, the performances under high pressure of 2 atm and the stability of sealed chamber with an $^{241}$Am source have been studied, and the relationship between the shape of Alpha particle spectra measured with gas gain and gas pressure has been studied and analyzed in detail.
The 8 groups of relative gas gain versus working voltage of THGEM under 1.3 $\sim$ 2.0 atm expressed by weighting filed $E/P$ are normalized, being consistent with theory.

The stability of long-term gas pressure has been monitored.
The pressure changes from 2.0 atm to 1.9 atm with the gas gain rising $\sim11.4\%$.
This tendency has been compared with Monte Carlo simulation on energy deposition without gas gain involved.
This is helpful for further study and application of the performance of sealed chambers.

For the next step, to answer whether the gain fall with time is caused by the outgas or some other factors, Some special instruments, for example Residual Gas Analyzer (RGA) will be used to analyze the change of the components of the working gas. In addition, to restrain the outgas of the chamber, some actions could be taken, such as baking the chamber under vacuum condition and changing the THGEM substrate with lower outgas effect.

Both the sealed and high gas pressure THGEM detectors would have broad scale applications with different kinds of particles.
Based on the results of this experiment, Gas Avalanche Photomultiplier to detect UV or visible light, Time Projection Chamber used in collider physics and other application in the field using high pressure and sealed chamber should be explored in the future.

\

\end{multicols}

\vspace{15mm}

\begin{multicols}{2}

\subsection*{Appendix A}

\end{multicols}
\ruleup
\begin{center}
\includegraphics[width=15cm]{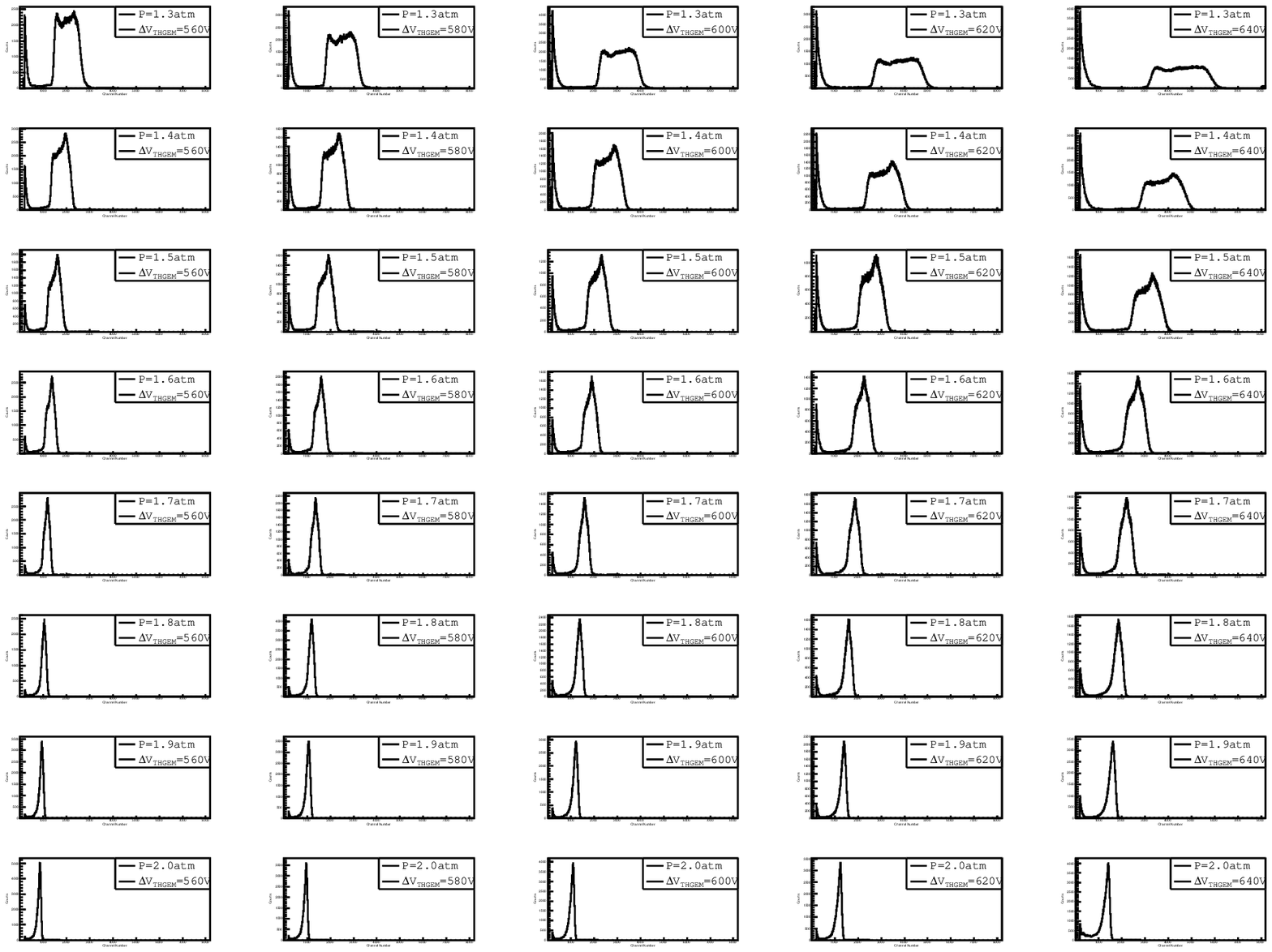}
\figcaption{\label{fig2} The MCA spectrum of Alpha particles for different values of gas pressure.}
\end{center}
\ruledown

\begin{multicols}{2}

\end{multicols}

\vspace{-1mm}
\centerline{\rule{80mm}{0.1pt}}
\vspace{2mm}

\begin{multicols}{2}

\end{multicols}

\clearpage

\begin{thebibliography}{90}

\vspace{3mm}



\bibitem{THGEM1}R. Chechik et al, Nucl. Instrum. Meth. A, {\bf 535}: 303---308 (2004)

\bibitem{THGEM2}C. Shalem et al, Nucl. Instrum. Meth. A, {\bf 558}: 475---489 (2006)

\bibitem{THGEM3}A. Breskin et al, Nucl. Instrum. Meth. A, {\bf 598}: 107---111 (2009)

\bibitem{hb}H. B. LIU et al, Nucl. Instrum. Meth. A, {\bf 659}: 237---241 (2011)

\bibitem{thinnerTHGEM}H. B. LIU et al, JINST, {\bf 7}: C06001 (2012)

\bibitem{qian}Q. LIU et al, JINST, {\bf 8}: C11008 (2013)

\bibitem{Bressan}A. Bressan et al, Nucl. Instrum. Meth. A, {\bf 423}: 119---124 (1999)

\bibitem{Buzulutskov}A. Buzulutskov et al, Nucl. Instrum. Meth. A, {\bf 433}: 471---475 (1999)

\bibitem{korff}A. S. Korff, \emph{Electron and Nuclear Counters} (New York: VanNostrand, 1955)

\bibitem{sauli}F. Sauli, \emph{PRINCIPLES OF OPERATION OF MULTIWIRE PROPORTIONAL AND DRIFT CHAMBERS}, CERN-77-09 (1977)

\bibitem{Charpak}G. Charpak et al, JINST, {\bf 3}: P02006 (2008)

\bibitem{nine}H. B. Xue et al, Proceedings of the 4th National Conference on Nuclear Monitoring (1999) (in Chinese)

\bibitem{Paredes}B. L. Paredes et al, JINST, {\bf 10}: P07017 (2015)

\bibitem{Saito}K. Saito et al, IEEE Transactions on Nuclear Science, {\bf 49}(4): 1674---1680 (2002)

\bibitem{Kim}D. Y. Kim et al, Journal of the Korean Physical Society, {\bf 68}: 1060---1068 (2016)

\bibitem{geant}S. Agostinelli et al, Nucl. Instrum. Meth. A, {\bf 506}: 250---303 (2003)

\bibitem{Amaro}F. D. Amaro et al, Nucl. Instrum. Meth. A, {\bf 579}: 62---66 (2007)

\bibitem{Dong}J. Dong et al, Chinese Phys. B, {\bf 18}: 4299---4233 (2009)

\bibitem{zhou}X. K. Zhou et al, Chin. Phys. Lett, {\bf 31}(3): 032901 (2014)

\bibitem{csi}H. B. LIU et al, CPC(HEP $\And$ NP), {\bf 35}(4): 363---367 (2011)

\bibitem{Ropelewski}F. Sauli et al, Nucl. Instrum. Meth. A, {\bf 560}: 269---277 (2006)

\bibitem{Alexeev}M. Alexeev et al, JINST, {\bf 8}: P01021 (2013)

\bibitem{wang}B. L. Wang et al, Chin. Phys. Lett, {\bf 31}(12): 122901 (2014)

\bibitem{alpha}Data for the $^{241}$Am source. \url{https://en.wikipedia.org/wiki/Americium#cite_note-92}.

\bibitem{sitar}B. Sitar et al, \emph{Ionization Measurements in High Energy Physics} (Berlin: Springer, 1993), p. 89



\end{thebibliography}
\end{document}